\documentclass[showpacs,aps,twocolumn
]{revtex4}
\usepackage{graphicx}
\usepackage{amsfonts}
\usepackage{amsmath}
\usepackage{amssymb}
\usepackage{multirow}
\usepackage{braket}
\usepackage[usenames,dvipsnames]{color}
\usepackage[english]{babel}
\usepackage[cp1251]{inputenc}   

\begin{document}
\bibliographystyle{apsrev}

\title{Information temperature as a measure of DNA's and texts complexity} 

\author{ O.~V.~Usatenko}

\affiliation{A. Ya. Usikov Institute for Radiophysics and
Electronics Ukrainian Academy of Science, 12 Proskura Street, 61805
Kharkov, Ukraine}

\begin{abstract}
C. Shannon introduced the notion of entropy for random sequences. What about their temperature? After discussing some methods for introducing information temperature (IT) for binary random stationary ergodic sequence, we suggest using IT as a characteristic of complexity and intelligence of an agent capable of writing or generating meaningful texts. In particular, we discuss the question of whether the temperature can characterize the academic level of the text, or serve as an
indicator of the quality of brain activity of the text author.
\end{abstract}

\maketitle


\section{Introduction}

The description and understanding of the complexity of natural and artificial dynamical systems remains an open problem in science. In connection with the rapid development of neural networks and artificial intelligence, in order to assess progress in the field of AI, the question inevitably arises of the possibility of comparing various models and devices that simulate intelligence and their comparison with the real human intelligence.

Quantitative evaluation of artificial intelligence faces currently several important challenges. A fundamental problem in artificial intelligence is that nobody really knows what intelligence is. There is no standard definition of what exactly constitutes intelligence. We do not have unified models of an artificially intelligent system. Others believe that intelligence encompasses a range of aptitudes, skills, and talents. Despite all these complexities, artificial intelligence systems are improving, and currently well-known examples include self-driving cars, image recognizers and classifiers, translators, chess and Go game simulations, and many others.

Human activity is so multifaceted that it is hardly possible to invent a universal measure of his intellect. It is difficult to compare who is the more gifted and brilliant writer or scientist, designer or artist, actor or director.

On this path, the question inevitably arises about intelligence in what kind of activity? We will consider intelligence as capability to produce meaningful texts.

Currently, we can present and discuss briefly several way defining the level of intelligence activity. We give here a number of well known informal definitions of human intelligence that have been earlier given.

The classic approach to determining whether a machine is intelligent is the so called Turing test~\cite{Turing,Living} which has been extensively debated over the last 50 years. The idea is that if a machine can successfully imitate a human, then this machine is AI. A computer program called Eugene Goostman, which simulates a 13-year-old Ukrainian boy, is said to have passed the Turing test at an event organized by the University of Reading.

IQ, or intelligence quotient test~\cite{IQ}, is a measure of ability to reason and solve problems. It essentially reflects how well we does on a specific test as compared to other people of some age group. Tests for the ability of artificial intelligence are developing in parallel with the understanding of what it really is.

Several authors \cite{Chaitin82,Chaitin92,Orallo,CaDoOral} have suggested the relevance of compression to intelligence, especially the inductive learning part of intelligence.

Legg and Hutter~\cite{LeggHutter05,LeggHutter07}  gave ten definitions of human intelligence and formulated its precise definition to the term \emph{Universal} Intelligence.

Among other papers let us mention \emph{objective} definition of intelligence~\cite{McCarthy} and integrated information theory \cite{Tononi} which has established itself as one of the leading theories for the study of consciousness. Tegmark~\cite{Tegmark} suggests that there are as many types of consciousness as there are types of matter.

In this work, we propose a new quantitative test for comparing the intellectual capabilities of the AI simulator and living creatures using information temperature.
Our measure of intelligence is aimed at finding out the ability to write meaningful texts. We will characterize the complexity of written text by its informational temperature.

The structure of the rest of the article is as follows. In Section \ref{GenDef}
we provide brief description of the binary exactly solvable Markov chain of order $N\leqslant 2$ with step-wise memory for which we introduce the information temperature. In Section~\ref{Entropy} we present two simplest examples of the application of the concept of information temperature: the dependence of the entropy and complexity of the system on the parameter of ordering of the system. Section IV contain discussion of the use of the informational temperature to characterize the informational and intellectual complexity of people and artificial intelligence systems assuming that the latter, both human and AI, are represented by written/generated texts.

\section{Information temperature}\label{GenDef}
In this section we introduce briefly a definition of
the information temperature. More detailed description is contained in paper~\cite{UsMPYa}.

One of methods introducing the informational temperature (equivalent correspondence (EC)) is based on the fact that every binary Markov chain is equivalent to some binary two-sided chain \cite{AMUYa} which, in its turn, can be viewed as the Ising sequence at a fixed temperature where the probability of a configuration is given by the Boltzmann distribution.

The second method makes use of  the traditional entropy based  thermodynamic definition of temperature with direct calculation of  the block entropy and energy of Markov chain in the pair-interaction  approximation.

Both approaches give the same result for the case of nearest neighbor spin/symbol interaction but the method of correspondence of Markov and Ising chains becomes very cumbersome for longer range of correlations, $N\geqslant 3$.
Note that while the first way of the temperature introduction is quite
natural, the second one can be interpreted as heuristic or axiomatic. We dive here  two methods of introducing information temperature in order to show the validity and effectiveness of the method of introducing fictitious energy.

\subsection{Temperature from equivalence}\label{Equiv}
Consider an infinite random stationary ergodic sequence
$\mathbb{S}$  of symbols-numbers $a_{i}$,
\begin{equation}
\label{RanSeq} \mathbb{S}= ..., a_{0}, a_{1},a_{2},...
\end{equation}
taken from the binary alphabet $a_{i}\in \mathcal{A}$:
\begin{equation}\label{alph}
 \mathcal{A}=\{0,1\},\,\, \,\, i \in
\mathbb{Z} = \{...,-1,0,1,2...\}.
\end{equation}
A higher number
of states can be handled by binary coding.
We use the notation $a_i$ to indicate a position $i$ of the symbol
$a$ in the chain and the unified notation $\alpha^k$ to stress the
value of the symbol $a\in \mathcal{A}$. 

We suppose that the symbolic sequence $\mathbb{S}$ is a
\textit{high-order Markov chain}. The sequence $\mathbb{S}$ is a
Markov chain if it possesses the following property: the probability
of symbol~$a_i$ to have a certain value $\alpha^k \in \mathcal{A}$
under the condition that {\emph{all}} previous symbols are fixed
depends only on $N$ previous symbols,
\begin{eqnarray}\label{def_mark}
&& P(a_i=\alpha^k|\ldots,a_{i-2},a_{i-1})\\[6pt]
&&=P(a_i=\alpha^k|a_{i-N},\ldots,a_{i-2},a_{i-1}).\nonumber
\end{eqnarray}

For a \emph{two-sided random chain} the conditional probability that symbol~$a_i$ is equal to unity, under condition
that the \emph{rest} of symbols in the chain are fixed, can
be presented of the form~\cite{AMUYa},
\begin{equation} \label{2}
P(a_i=1|A_i^-,A_i^+)
\end{equation}
\begin{equation}
=\displaystyle\frac{P(a_i=1,A_i^+|A_i^-)}
{P(a_i=1,A_i^+|A_i^-)+P(a_i=0,A_i^+|A_i^-)},\nonumber
\end{equation}
where $A_i^-$ and $A_i^+$ are the semi-infinite words,
$(\ldots,a_{i-2},a_{i-1})$ and $(a_{i+1},a_{i+2},\ldots)$,
surrounding symbol $a_i$ respectively are,
\begin{equation}
\underbrace{\ldots,a_{i-2},a_{i-1}}_{A_i^-},a_i,\underbrace{a_{i+1},a_{i+2},\ldots}_{A_i^+}.
\end{equation}
Here the two-sided conditional probability $P(a_i=1|A_i^-,A_i^+)$ is expressed
by means of the Markov-like probability functions $P(a_i
,A_i^+|A_i^-)$. These probabilities containing semi-infinite
one-sided words can be expressed in terms of the conditional
probability function of the Markov chain.

For the stationary Markov chain, the probability $b(a_{1}a_{2}\dots
a_{N})$ of occurring a certain word $(a_{1},a_{2},\dots ,a_{N})$
satisfies the condition of compatibility for the Chapman-Kolmogorov
equation (see, for example, Ref.~\cite{gar}):
\[
b(a_{1}\dots a_{N})
\]
\begin{equation}
=\sum_{a=0,1}b(aa_{1}\dots a_{N-1})P(a_{N}\mid a,a_{1},\dots
,a_{N-1}).  \label{10}
\end{equation}
In works~\cite{UYa} and~\cite{UYaKM}, we have introduced the model
Markov chain for which the conditional probability $p_{k}$ of
occurring the symbol ``0'' after the $N$-symbol words containing $k$
unities, e.g., after the word
$\underbrace{(11...1}_{k}\;\underbrace{00...0}_{N-k})$, is given by
the following expression:
\[
p_{k}=P(a_{N+1}=0\mid \underbrace{11\dots
1}_{k}\;\underbrace{00\dots 0} _{N-k})
\]
\begin{equation}
=\frac{1}{2}+\mu (1-\frac{2k}{N}).  \label{1}
\end{equation}

For the ordinary one-step Markov chain with nearest-neighbor interaction using Eq.~\eqref{1} with $N=1$,  we easily obtain the conditional probabilities
\begin{eqnarray}\label{111}
&& P(a_i=1,a_{i+1}=1|a_{i-1}=1)=\left(\frac{1}{2}+\mu\right)^2,\\[6pt]
&&P(a_i=0,a_{i+1}=1|a_{i-1}=1)=\left(\frac{1}{2}-\mu\right)^2.\nonumber
\end{eqnarray}

Considering the equivalence of Markov chain to the
two-sided random sequence, we apply
formula~\eqref{2} to the word $(1,1,1)$:
%
\begin{eqnarray}\label{TwoOne}
&& P(a_i=1|a_{i-1},a_{i+1})\\[6pt]
&&=\frac{P(a_i=1,a_{i+1}|a_{i-1})}{P(a_i=1,a_{i+1}|a_{i-1})+
P(a_i=0,a_{i+1}|a_{i-1})}.\nonumber
\end{eqnarray}

The Boltzmann distribution for the corresponding Ising model gives,
\begin{eqnarray}\label{P Ising}
&& P(a_i=1|a_{i-1}=1,a_{i+1}=1)=P(\uparrow\uparrow\uparrow)\nonumber \\[6pt]
&&=\frac{\exp(2\varepsilon/T)}{\exp(2\varepsilon/T)+\exp(-
2\varepsilon/T)}.
\end{eqnarray}
The energy of the spins interaction are $\varepsilon_{\uparrow
\uparrow }=\varepsilon_{\downarrow\downarrow}=-\varepsilon$,
$\varepsilon_{\uparrow \downarrow
}=\varepsilon_{\downarrow\uparrow}=\varepsilon>0.$
Using Eqs.~\eqref{111} - \eqref{P Ising}, we have
\begin{equation}
  \mu = \frac{1}{2} \tanh {\left( \frac{1}{\tau} \right)}, \quad\quad
  \beta=\frac{1}{\tau} = \frac{1}{2} \ln \frac{1+2\mu}{1-2\mu}.
 \label{MuVsT}
\end{equation}
Here we have introduced the informational temperature $\tau=
T/\varepsilon$ of the ordinary, one-step, Markov chain and $\beta=\varepsilon/T$
is the inverse info-temperature. So, for $\tau \rightarrow {\pm\infty}$  we have $\varepsilon/T \simeq 2\mu
\rightarrow 0$, and $\tau \rightarrow \pm 0$ when $\mu \rightarrow
\pm {1/2}$. The negative values of $\tau$
describes an anti-ferromagnetic ordering of spines or symbols\,
"0"\, and "1". We can say that $\tau$ is the temperature $T$
measured in unites of  energy $\varepsilon$, $\tau={T}/
\varepsilon$. Note that the informational temperature depends only on the Markov chain parameter $\mu$, and not on the introduced fictitious energy $\epsilon$.

\subsection{Temperature from the entropy}\label{EntrTemp}
In this section we present another method of introducing the information temperature based on thermodynamic definition of temperature with direct calculation of  the block entropy and energy of Markov chain in the pair-interaction  approximation. The blocks are chosen to be $(N+1)$-length subsequences of symbols occurring  due to the fact that they encompass all the information about the structure of the $N$-step Markov chain. The probabilities of the blocks/words  are defined by the Chapman-Kolmogorov equation.

Using Eqs.~\eqref{10} and \eqref{1} for $N=2$, we find the
probabilities of 2-symbols words
and 3-points probabilities of symbols $000$ and $111$ occurring.
Introducing fictive energies of different spin configurations:
\begin{equation}\label{000111}
  \varepsilon(\downarrow\downarrow\downarrow)=\varepsilon(\uparrow\uparrow\uparrow)
    =-2\varepsilon_1-\varepsilon_2,
\end{equation}
\begin{equation}\label{010101}
    \varepsilon(\downarrow \uparrow \downarrow)=\varepsilon(\uparrow \downarrow\uparrow)
    =2\varepsilon_1-\varepsilon_2,
\end{equation}
\begin{equation}\label{allothers}
    \varepsilon(\downarrow \downarrow \uparrow)=...
    =\varepsilon(\uparrow \uparrow\downarrow)=\varepsilon_2,
\end{equation}
we calculate the energy %
\begin{equation}\label{2Temperature}
    E_3 = -(2\varepsilon_1+\varepsilon_2)\frac{\mu}{1-\mu},
\end{equation}
and the entropy per two bonds of random elements:
\begin{equation}\label{2S}
  H_3 = -2 P_{000}\ln P_{000}-6 P_{001}\ln P_{001}.
\end{equation}

Here we use the well-known definition of entropy  of subsequence of
symbols of length $L$, see,
e.g., Refs.~\cite{Cover,Ebeling},
\begin{eqnarray} \label{entro_block}
H_{L}=-\sum_{a_{1},...,a_{L} \in \mathcal{A}} P(a_{1}^{L})\log_{2}
P(a_{1}^{L}).
\end{eqnarray}
Here $P(a_{1}^{L}) =P(a_{1},\ldots,a_{L})$ is the probability to
find the $L$-word $a_{1}^{L}$ in the sequence and $L=3$ in the considered here case.

Calculating derivative $dH_3/dE$ as $(dH_3/d\mu)/(dE_3 /d\mu)$, we
obtain the following result:
\begin{equation}\label{2tau}
    \frac{1}{\tau}=\frac{1}{4} \ln
    \frac{1+2\mu}{1-2\mu},
    \end{equation}
where
\begin{equation}\label{AvEps}
    \frac{1}{\tau}=\frac{\langle\, \varepsilon\rangle }{T}, \quad\, \langle\, \varepsilon\rangle=
    \frac{2\,\varepsilon_1+\varepsilon_2}{3}.
\end{equation}
Emerged here average quantity $\langle\, \varepsilon\rangle$ can be
treated formally as the average (fictive) energy of symbols
interaction and is completely different from the energy $E_3$, Eq.~\eqref{2Temperature}. At the same time, $\langle\, \varepsilon\rangle$ can be considered as
the unity of measurement of the temperature $T$. Putting in Eq.~\eqref{AvEps}
$\varepsilon_1=\varepsilon_2=1$, we obtain natural unity of the
information temperature measure, $\tau=T$.

Let us note that the result presented by Eq.~\eqref{2tau} is obtained
without any reference to the spin chain. Another important point of
the above consideration is independence of
Eq.~\eqref{2tau} from the introduced arbitrary energies
$\varepsilon_1$ and $\varepsilon_2$ of the symbols interactions. Thus, despite the fact that the method of matching random sequences seems to be more natural and consistent, it turns out to be more difficult and restrictive in comparison with the entropy method.  Note that using the method of this section we can recover the result obtained above for the one-step Markov chain, Eq.~\eqref{MuVsT} and vis-versa the result Eq.~\eqref{2tau} can be obtained by the method of equivalence of the previous suction.

\section{Temperature characterizes entropy and complexity}\label{Entropy}

Thus, we have above presented our earlier finding~\cite{UsMPYa} on the informational temperature which should be considered as  macroscopic statistical characteristic of stationary ergodic random sequences. The temperature is one of the most important macroscopic parameters in statistical physics, and we believe that the concept of informational temperature should also make deep sense in relation to random sequences. In this section of the work, we present two simplest examples of the application of the concept of information temperature. Earlier, in the works of a number of authors~\cite{Hogg,Li,Lopez}, the question of the dependence of the entropy and complexity of the system on the parameter of ordering of the system was discussed. This parameter did not have a clear and unambiguous name and definition. It was called either ``degree of disorder'' of the system~\cite{Hogg}, or the parameter ``order -- disorder''~\cite{Li}, or a system parameter ranging
``from ideal crystal to ideal gas''~\cite{Lopez}. If earlier the answer to the question on what parameters the entropy depends could be interpreted ambiguously, now we give the natural answer to this question -- it is the informational temperature.

\begin{figure}[h!]
\begin{centering}
\scalebox{0.3}[0.3]{\includegraphics{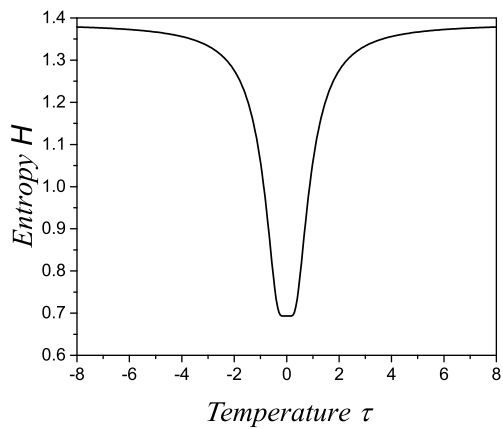}}
\caption{The entropy $H_2(\tau)$ for the 1-step Markov chain. The positive values of $\tau$ describe persistent ferromagnetic ordering of the random stationary ergodic sequence $\mathbb{S}$, the negative ones correspond to anti-persistent anti-ferromagnetic case.} \label{Fig1}
\end{centering}
\end{figure}

The entropy per one bond $H_2$ for the ordinary one-step Markov chain considered in Section~\ref{Equiv} can be found using the method of Section \ref{EntrTemp}
or using the result of paper~\cite{Li}:
\begin{equation}
  H_2 = -2 P_{00}\ln P_{00}-2 P_{01}\ln P_{01}= \ln 2 \nonumber
  \end{equation}
\begin{equation}\label{S}
 -(1/2+\mu) \ln (1/2+\mu)-(1/2-\mu) \ln (1/2-\mu).
\end{equation}
Here $P_{ij}$ are the probabilities of the
words of the length two  occurring:
\begin{equation}\label{00}
P_{00}=P_{11}=(1/2+\mu)/2,
\end{equation}
\begin{equation}\label{01}
    P_{01}=P_{10}=(1/2-\mu)/2,
\end{equation}
and the parameters $\mu$ is defined by Eq.~\eqref{MuVsT}.
As a result we get the entropy as a function of info-temperature,
\begin{eqnarray}\label{H2}
H_2&=&(\ln 2 + \ln(1+\exp(-2\beta))/(1+\exp(-2\beta))  \nonumber \\[1pt]
&+& \ln(1+\exp(2\beta))/(1+\exp(2\beta))).
\end{eqnarray}
This is a very natural expression that approaches its minimum and maximum values as $\tau=1/\beta$ approaches zero or infinity, respectively. This result is presented in Fig.~\ref{Fig1}.

Earlier it was pointed out~\cite{Hogg,Li} that the algorithmic (entropic) complexity is not
corresponds to our intuitive understanding of complexity. In this regard, the paper~\cite{Lopez} proposes a definition that discriminates complexity in the region of high disorder or, in our terms, when $\tau \rightarrow \infty$. This is simply the interplay between the information stored in the system and its disequilibrium.
In our case, this definition has the following form:
\begin{eqnarray}
C = h D, \quad\quad D=\sum_{ij ={0,1}} \left(P_{ij} -\frac{1}{4}\right).
\end{eqnarray}
Here $h =H_2-H_1$ is the Shannon entropy, or the entropy per symbol,
 $H_1=\ln 2$, $D$ is the so called disequilibrium factor, is equal in our case to the value $D=\mu^2$, obtained with using Eqs.~\eqref{00} and \eqref{01}. The final result for the complexity expressed via the informational temperature $\tau$ is
\begin{eqnarray}\label{h D}
&&C(\tau)= \tanh^2(1/\tau)\\[6pt]
&\times&\left[ \frac{\ln(1+\exp(-2/\tau))}{1+\exp(-2/\tau)} + \frac{\ln(1+\exp(2/\tau))}{1+\exp(2/\tau)}\right].\nonumber
\end{eqnarray}

\begin{figure}[h!]
\begin{centering}
\scalebox{0.3}[0.3]{\includegraphics{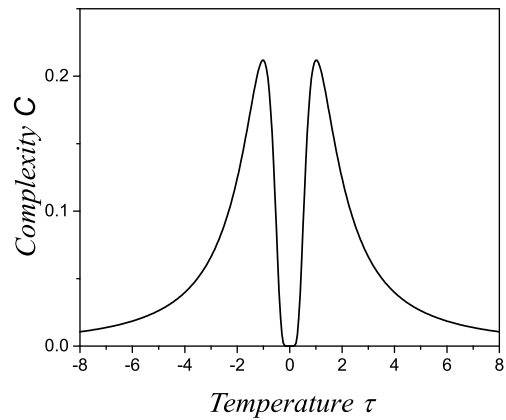}}
\caption{The dependence of complexity $C(\tau)$ on the informational temperature for the 1-step Markov chain.} \label{Fig2}
\end{centering}
\end{figure}

This result is presented by the  curve in Fig.~\ref{Fig2}.

\section{Temperature is a parameter of a text complexity}
Thus, in the previous section, we have presented two simplest examples of the application of the concept of informational temperature. We have seen that both entropy and information complexity can be considered as temperature dependent functions. Consequently, the temperature itself characterizes the measure of the complexity of the system. We believe that the importance of the concept of informational temperature is not limited to these examples and has a much wider scope.

We suppose that the informational temperature can be used to characterize the informational and intellectual complexity of people and artificial intelligence systems assuming that the latter, both human and AI, are represented by written/generated texts. However, we note the insufficiency of the obtained expressions for their use in applied purposes, since the memory scale can reach several thousand units, while the results obtained are valid only for a simple one-step Markov chain.

%

D. Tononi ~\cite{Tononi} indicates the possibility of
characterizing a neural network by some $\Phi$ function.
This function is not clearly defined and difficult for calculations. Can a neural network be characterized by the temperature of the text that the neural network is able to generate? We suppose that the answer to this question is positive.

\begin{acknowledgments}
The author of this work apologizes to readers for the fact that the work is published unfinished. The author of the work is not sure about the possibility of its completion, as he is in a place under continuous artillery and bombing.
\end{acknowledgments}

\end{document}